\documentclass[a4paper,twoside,10pt]{article}
\usepackage{graphicx,publaob}
\usepackage[table,x11names]{xcolor}

\def\papertype{Contributed paper}
\begin{document}

\title{INVESTIGATING THE RADIAL ACCELERATION RELATION IN EARLY-TYPE GALAXIES USING THE JEANS ANALYSIS}

\authors{M. B\'{i}lek$^{1,~2,~3,~4}$ and S. Samurovi\'c$^5$}

\address{$^1$Astronomical Institute, Czech Academy of Sciences, Bo\v cn\'{\i} II 1401/1a, CZ-14100 Prague, Czech Republic,}

\address{$^2$Faculty of Mathematics and Physics, Charles University in Prague, Ke Karlovu 3, CZ-121 16 Prague, Czech Republic,}
\address{$^3$Helmholtz-Institut f\" ur Strahlen und Kernphysik, Nussallee 14-16, 53115 Bonn, Germany,}

\address{$^4$Charles University in Prague, Faculty of Mathematics and Physics, Astronomical Institute, V Hole\v sovi\v ck\' ach 2, 180 00 Prague 8, Czech Republic,}
\Email{michal.bilek}{asu.cas}{cz}

 \address{$^5$Astronomical Observatory, Volgina 7, 11060 Belgrade, Serbia}
\Email{srdjan}{aob.bg.ac}{rs}

\markboth{RAR IN ETGs}{M. B\'ILEK and S. SAMUROVI\'C}

\abstract{Investigating the gravitational field in the early-type galaxies (ETGs, i.e. ellipticals and lenticulars) up to large radii is observationally difficult. It is questionable how the radial acceleration (RAR) in the ETGs looks like, i.e. the relation between the dynamically inferred gravitational acceleration and the acceleration expected from the distribution of the visible matter. This relation is nearly universal for the spiral galaxies, in agreement with the MOND modified dynamics paradigm. In this contribution, we investigate a sample of 15 ETGs. We extract their full kinematic profiles out to several effective radii from their globular cluster systems and estimate their gravitational field using the Jeans equation. We parametrize the gravitational field by that produced by the stars and a Navarro-Frenk-White DM halo. We find that only 4-5 of our ETGs follow the RAR for the spiral galaxies.  All these galaxies are fast rotators, have disky isophotes, appear mostly very elongated and the have bluest colors in our sample. This suggests that they might be spiral galaxies which lost their gas. Our galaxies deviating from the RAR for the spirals either disprove MOND, contain unobserved matter, or indicate a flaw in the method.
}

\section{INTRODUCTION}

The missing mass problem is far from being solved. While the usual explanation of the dynamical discrepancies is the presence of the dark matter (DM), the theories suggesting modifications of the standard laws of physics are viable. Among these theories, MOND (Milgrom 1983) is particularly popular and successful. It states, roughly speaking, that the actual gravitational acceleration, $a$, is a function of the gravitational acceleration calculated in the Newtonian way from the distribution of the visible matter, $a_\mathrm{N}$. The acceleration $a$ differs from $a_\mathrm{N}$ only when  $a_\mathrm{N}$ is lower than about $a_0=1.35^{+0.28}_{-0.42}\times 10^{-8}$ ${\rm cm\,s^{-2}}$ (Famaey et al.~2007). This functional dependence has been tested many times (see the review by Famaey \& McGaugh, 2012), mainly in the late-type galaxies (LTGs, i.e. the spiral galaxies).  Indeed, observations show that the relation between $a$ and $a_\mathrm{N}$ (the acceleration relation, RAR) has a very low scatter for LTGs of a wide variety of sizes, masses, gas fractions and environments. According to the extensive study by  McGaugh et al. (2016), the best currently available data for the LTGs are consistent with no intrinsic scatter in the RAR, as expected by MOND. The tightness of the RAR for the LTGs remains unexplained with the DM hypothesis.

Investigations of the RAR in the early-type galaxies (ETGs, i.e. the elliptical and lenticular galaxies) is much less advanced. These galaxies typically lack the disks of rotating gas that enable measuring the rotational velocity, and hence the acceleration $a$, in the LTGs up to large radii. There are ways to investigate the gravitational fields near the ETG centers, such as from the velocity dispersion profiles of stars or from the temperature and luminosity profiles of the hot gas coronas. These measurements are however available only in the regions where the acceleration $a$ is relatively high compared to the boundary acceleration $a_0$. If we want to probe the weak gravitational field at the outskirts of the ETGs, we can rely, more or less, only on the Jeans analysis of  the tracers such as planetary nebulae and globular clusters (GCs) (see, e.g., Samurovi\'c 2007). With this method, we are looking for an analytical profile of the gravitational field of the galaxy that can explain the observed velocity dispersion profiles of the GC systems. Such investigations require measuring radial velocities of many GCs which have to be obtained by very time consuming observations at large telescopes. This is the reason why the number of the ETGs studied using the Jeans analysis slowly increases.

Here we study using the Jeans analysis the gravitational field profiles up to 5 effective radii ($R_\mathrm{e}$) for 15 ETGs, which is close to the largest sample  possible today. We use these profiles to investigate the RAR in ETGs in unprecedented detail. We find that only 4 or 5 of our ETGs follow the same RAR as the LTGs.

\section{OBSERVATIONAL DATA}

The ETGs in our sample are objects with a wide range of luminosities, morphologies and environments (field, group and cluster galaxies). We use observational data from several sources that will  be listed in our prepared paper (most of them come from Pota et al., 2013). We work with a {\it total} sample of GCs for each galaxy (not splitting into red and blue sub-populations) in order to have more clusters per bin because our goal is to determine as accurately as possible the velocity dispersion and departures of the GC radial velocity distribution from a Gaussian. Table~\ref{tab:1} lists our objects and their properties.

 \section{OUR METHOD}

\subsection{Dynamical models}

For all 15 ETGs in our sample, we solve the Jeans equation (e.g.~Binney \& Tremaine 2008) in the spherical approximation
\begin{equation}
{{\rm d}\sigma_r^2 \over {\rm d}r}
+  (\alpha+2\beta){\sigma_r^2 \over r} = a + {v_{{\rm rot}}^2\over r},
\label{eqn:Jeans1}
\end{equation}
where $\sigma_r$ is the radial velocity dispersion of the GCs and $\alpha={\rm d}\ln \nu /{\rm d}\ln r$ is the slope of their projected number density $\nu$. We analyze the isotropic case ($\beta=0$), the 
tangentiallly dominated case ($\beta=-0.2$) and a radially dominated case  ($\beta_{\rm lit}>0$). More details are available in Samurovi\'c (2014, hereafter S14). 
We model the gravitational potential of our galaxies as $\phi = \phi_*+ \phi_\mathrm{NFW}$, where $ \phi_*$ is the Newtonian gravitational potential generated by the stars observed in the galaxy and  $\phi_\mathrm{NFW}$ is the Newtonian potential of an NFW DM halo  (Navarro, Frank \& White 1997). There are three free parameters: the stellar mass-to-light ratio and the characteristic radius and density of the halo. We substitute $-\nabla\phi = a$ to Eq.~\ref{eqn:Jeans1} and find the free parameters so that the deviation of the observed velocity dispersion profile from the velocity dispersion profile given by Eq.~\ref{eqn:Jeans1} is minimized. The minimizing potential is adopted as the real gravitational potential of the galaxy.

\subsection{Baryonic models}

In order to obtain the acceleration $a_\mathrm{N}$, we had to derive the stellar mass-to-light ratio, $M/L_*$, of every galaxy. When it is known and the galaxy is assumed to be spherical, then  the acceleration $a_\mathrm{N}$ can be calculated from the observed luminosity distribution. We used several methods to derive $M/L_*$: 1) from the color index of the galaxy and stellar population synthesis (SPS) models, 2) from the dynamical models as the dynamical mass cumulated under $0.5\,R_\mathrm{e}$ divided by the cumulated luminosity under the same radius; this is based on the fact that no DM is usually needed in ETG centers, 3) similarly, we estimate $M/L_*$ from the dynamical mass cumulated below $1\,R_\mathrm{e}$.

\section{RESULTS AND DISCUSSION}

\begin{figure}
         \begin{center}              
         \includegraphics{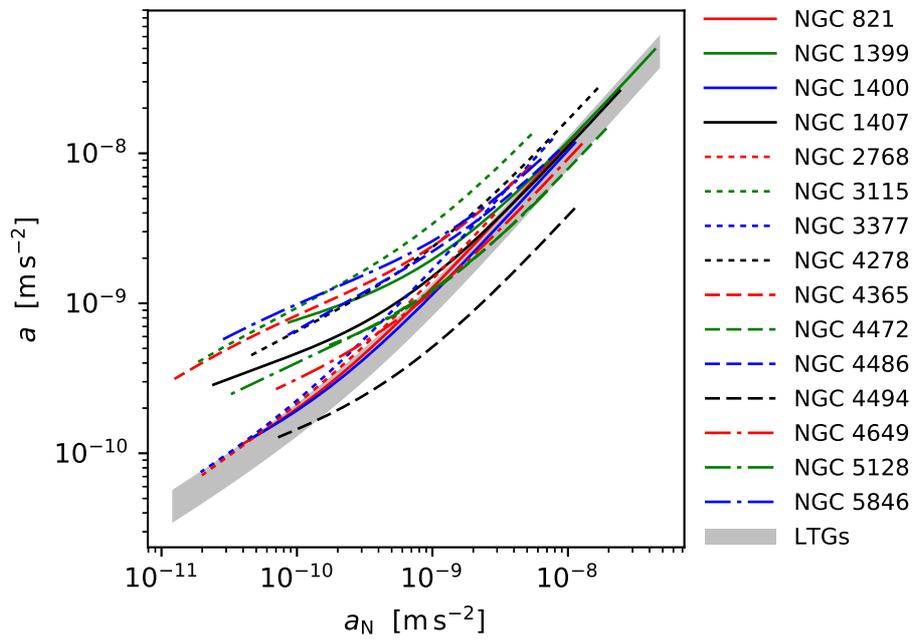}
        \caption{The radial acceleration relations of our 15 ETGs. Here the $M/L_*$ from an SPS model was used. The RARs of the LTGs lie mostly in the gray region (McGaugh et al. 2016).}\label{fig:1}
        \end{center}
\end{figure}

Figure~\ref{fig:1} shows the RARs found for the ETGs in our sample. The gray area indicates the common RAR for the 153 LTGs from McGaugh et al. (2016, their Eq.~4) where the vertical thickness corresponds to the $\pm 1\,\sigma$ average scatter (as given at their Fig.~3). We can see that only 4 of our 15 ETGs have their RAR roughly common with the RAR of the LTGs. They are NGC\,821, NGC\,1400, NGC\,2768 and NGC\,3377. This result does not change for any choice of the dynamical or baryonic models discussed above; possibly only NGC\,4494 could be added to the four common galaxies for another choice of $M/L_*$.

We note that our ETGs following the RAR of the LTGs resemble the LTGs also in other regards (see Table~\ref{tab:1}): their stellar content is supported more by ordered rotation than velocity dispersion, they have disky isophotes, they sometimes appear very flattened in projection, they have the bluest colors in our sample, and avoid galaxy clusters. These facts suggest that their formation history is related to the formation history of the LTGs. They could, for example be spiral galaxies which lost their gas by a starburst. We also note they can be modeled in MOND without additional DM,  see S14, Samurovi\'c et al. (2014) and also our forthcoming paper. On the other hand, massive ETGs, such as NGC~3115, NGC~4365 and NGC~5846 which deviate significantly from the RAR of the LTGs, need copious amounts of DM in their outer regions even in the MOND approach (see S14). 

One possibility is that our results exclude MOND. Given its previous success, including in the ETGs (e.g., Milgrom, 2012; 2013, Tian \& Ko, 2016; 2017), our results should be verified for our galaxies by an independent method and the reliability of the Jeans analysis should tested against numerical simulations.
It is possible to reconcile MOND with our results by supposing additional invisible matter in the galaxies. MOND is already known to require some DM in galaxy clusters. The most discussed candidates are sterile neutrinos (Angus 2007) and compact baryonic objects (Milgrom 2008). We note that the DM required in our galaxies might be connected with the yet undetected gas which is predicted to flow into galaxies to explain various observations (Sancisi et al. 2008). 

\bigskip
ACKNOWLEDGEMENTS. {This work was supported by the Ministry of Education, Science and Technological Development of the Republic of Serbia through project no.~176021, ``Visible and Invisible Matter in Nearby Galaxies: Theory and Observations''. The  work  of MB was supported by the European Commission through project BELISSIMA (BELgrade  Initiative  for  Space  Science,  Instrumentation  and  Modelling  in
Astrophysics,  call  FP7-REGPOT-2010-5,  contract  No.  256772). } 
 
\references

Angus, G.W., Shan, H.Y., Zhao, H.S., \& Famaey, B.: 2007,  \journal{Astrophys. J.}, \vol{654}, L13.
 
Binney, J.J. \& Tremaine, S., 2008:  Galactic Dynamics, 2nd Ed., Princeton University Press.

Famaey, B. et al.: 2007, \journal{Phys. Rev. D}, \vol{75}, 063002.

Famaey, B., \& McGaugh, S.S.: 2012, \journal{Living Reviews in Relativity}, \vol{15}, 10.
           
McGaugh, S.S., Lelli, F., \& Schombert, J.M.: 2016, \journal{Phys. Rev. Lett.}, \vol{117}, 201101.
           
Milgrom, M.: 1983,  \journal{Astrophys. J.}, \vol{270}, 365.
           
Milgrom, M.: 2008, \journal{New Astron. Rev.}, \vol{51}, 906.

Milgrom, M.: 2012, \journal{Phys. Rev. Lett.}, \vol{109}, 131101.

Milgrom, M.: 2013, \journal{Phys. Rev. Lett.}, \vol{111}, 041105. 
 
Navarro, J.F., Frenk, C.S. \& White, S.D.M.: 1997,  \journal{Astrophys. J.}, \vol{490}, 493.
           
Pota, V. et al.: 2013,  \journal{Mon. Not. R. Astron. Soc.},  \vol{428}, 389.
           
Samurovi\'c, S.: 2007, Dark Matter in Elliptical Galaxies, \journal{Publications of the Astronomical Observatory of Belgrade}, No. 81.
           
Samurovi\'c, S.: 2014, \journal{Astron. Astrophys.}, \vol{570}, A132, 29 pp. (S14).
           
Samurovi{\'c}, S., Vudragovi\'c, A., Jovanovi\'c, M. \& \'Cirkovi\'c M.M.: 2014, \journal{Serb. Astron. J.}, 188, 29.
	   
Sancisi, R., Fraternali, F., Oosterloo, T., \& van der Hulst, T.: 2008, \journal{Astron. Astrophys. Rev.}, \vol{15}, 189.

Tian, Y., \& Ko, C.-M.: 2016, \journal{Mon. Not. R. Astron. Soc.}, \vol{462}, 1092.

Tian, Y., \& Ko, C.-M.: 2017, \journal{Mon. Not. R. Astron. Soc.}, \vol{472}, 765.
	   
\endreferences

\begin{table}
\begin{tabular}{lcccccccc}
\hline
\noalign{\smallskip}
\multicolumn{1}{c}{Name} &
\multicolumn{1}{c}{D} &
\multicolumn{1}{c}{M$_B$} &
\multicolumn{1}{c}{Type} &
\multicolumn{1}{c}{Env} &
\multicolumn{1}{c}{Rot} &
\multicolumn{1}{c}{Iso} &
\multicolumn{1}{c}{$B-V$} \\ 
\multicolumn{1}{c}{(1)} &
\multicolumn{1}{c}{(2)} &
\multicolumn{1}{c}{(3)} &
\multicolumn{1}{c}{(4)} &
\multicolumn{1}{c}{(5)} &
\multicolumn{1}{c}{(6)} & 
\multicolumn{1}{c}{(7)} & 
\multicolumn{1}{c}{(8)}\\

\hline
\noalign{\smallskip}

\rowcolor{lightgray} NGC\,0821    &    23.4   &   -20.82	&  E6		 &   F	   &   f &  D &  0.87  \\
NGC\,1399    &   19.0   &  -20.81	& E1pec 	 &  C	   &  s & D & 0.93  \\
\rowcolor{lightgray} NGC\,1400    &    26.8   &   -20.35	&  S0/E0 	 &   G	   &   f &  0 &  0.89  \\
NGC\,1407    &   26.8   &  -21.49	& E0		 &  G	   &  s & 0 & 0.95  \\
\rowcolor{lightgray} NGC\,2768    &    21.8   &   -21.26	&  E6/S0$\_1/2$	 &   G	   &   f &  D &  0.91  \\
NGC\,3115    &    9.4   &  -19.94	& S0		 &  F	   &  f & D & 0.90  \\
\rowcolor{lightgray} NGC\,3377    &    10.9   &   -19.32	&  E6		 &   G	   &   f &  D &  0.82  \\
NGC\,4278    &   15.6   &  -19.50	& E1-2  	 &  G	   &  f & B & 0.90  \\
NGC\,4365    &   23.3   &  -21.00	& E3		 &  G	   &  s & B & 0.93  \\
NGC\,4472    &   16.7   &  -21.71	& E2		 &  C	   &  s & B & 0.93  \\
NGC\,4486    &   17.2   &  -22.05	& E0		 &  C	   &  s & B & 0.92  \\
\rowcolor{lightgray} NGC\,4494    &    16.6   &   -21.07	&  E1-E2 	 &   G	   &   f &  D &  0.85  \\
NGC\,4649    &   17.3   &  -21.59	& E2/S0 	 &  C	   &  f & B & 0.93  \\
NGC\,5128    &    3.8   &  -20.55	& S0pec(Epec)    &  G	   &  f & ? & 0.87  \\
NGC\,5846    &   24.2   &  -21.34	& E0		 &  G	   &  s & B & 0.94  \\

\noalign{\smallskip}
\hline
\noalign{\medskip}

\end{tabular}
\caption{NOTES: Columns are as follows (the references will be available in our forthcoming paper). (1): name of the galaxy. (2): distance to the galaxy in Mpc. (3):  B-band absolute magnitude.  (4): morphological type. (5): environment of each galaxy ("G" means that the galaxy belongs to a group, "F" is a field galaxy and "C" means that the galaxy belongs to a cluster). (6) central rotator type ("s" indicates slow and "f" indicates slow rotators). (7): isophotes ("D" are disky isophotes, "B" are boxy isophotes and "0" are pure ellipses) (8): corrected $B-V$ colors. The gray rows mark the objects following the RAR of the LTGs.}\label{tab:1}
\end{table}

\end{document}